\documentclass[pre,epsfig,twocolumn,showpacs,preprintnumbers,amssymb]{revtex4}
\usepackage{bm,graphicx,amsmath}

\begin{document}
\title{Persistent random walk on a
one-dimensional lattice with random asymmetric transmittances}

\author{Zeinab Sadjadi}
\author{MirFaez Miri}
\email{miri@iasbs.ac.ir} \affiliation{Institute for Advanced
Studies in Basic Sciences (IASBS), P. O. Box 45195-1159, Zanjan
45195, Iran}

\begin{abstract}
We study the persistent random walk of photons on a
one-dimensional lattice of random asymmetric transmittances. Each
site is characterized by its intensity transmittance $t$ ($t' \neq
t $) for photons moving to the right (left) direction.
Transmittances at different sites are assumed independent,
distributed according to a given probability density
$\mathcal{F}(t,t' )$. We use the effective medium approximation
and identify two classes of $\mathcal{F}(t,t' )$ which lead to the
normal diffusion of photons. Monte Carlo simulations confirm our
predictions. We mention that the metamaterial introduced by
Fedetov et al. [Nano Letters \textbf{7}, 1996 (2007)] can be used
to realize a lattice of random asymmetric transmittances.
\end{abstract}

\pacs{05.40.Fb, 05.60.-k, 02.50.Ey}

\maketitle
\section{Introduction}\label{intro}

Random walks in random environments is a field of continuous
research\ \cite{barry, weiss, Bouchaud, kehr}. Hopping conduction
of classical particles or excitations\ \cite{alex}, transport in
porous and fractured rocks\ \cite{sahimi1}, and diffusive
transport of light in disordered media\ \cite{sheng}, are a few
examples.

Random walks with {correlated} displacements figure in a multitude
of different problems. Among the correlated walks, the {persistent
random walk} introduced by F\"{u}rth\ \cite{furth} and Taylor
\cite{tay}, is possibly the simplest one to incorporate a form of
momentum in addition to random motion. In its basic realization on
a one-dimensional lattice, a persistent random walker possesses
constant probabilities for either taking a step in the same
direction as the immediately preceding one or for reversing its
motion\ \cite{barry,weiss,kehr}. Generalized persistent random
walk models are utilized in the description of polymers\
\cite{godoy}, chemotaxis\ \cite{chemotaxis}, general transport
mechanisms\ \cite{t0,www}, Landauer diffusion coefficient for a
one-dimensional solid\ \cite{55}, etc.

Recently, the persistent random walk model is used to study the
role of liquid films for diffusive transport of light in foams\
\cite{miriA, miriE}. Diffusing-wave spectroscopy experiments have
confirmed the photon diffusion in foams\ \cite{exp}. A relatively
dry foam consists of cells separated by thin liquid films\
\cite{Weaire1999}. Cells in a foam are much larger than the
wavelength of light, thus one can employ ray optics and follow a
light beam or photon as it is transmitted through the liquid films
with a probability $t$ called the intensity transmittance. This
naturally leads to a persistent random walk of the photons. In the
ordered honeycomb (Kelvin) foams, the one-dimensional persistent
walk arises when the photons move perpendicular to a cell edge
(face). Thin-film transmittance depends on the film thickness.
Films are not expected to have the same thickness. These
observations motivated us to consider persistent random walk on a
one-dimensional lattice of {\it random} transmittances\
\cite{miriE}. We assumed that transmittances at different sites
are independent random variables, distributed according to a given
probability density $f(t)$. Assuming that $<{1}/{t}>=\int_{0}^{1}
{f(t)}/{t} ~dt$ is finite, we validated the classical persistent
random walk with an effective transmittance $t_{eff}$, where
$1/t_{eff}=<1/t>$. We also investigated the transport on a line
with infinite $<1/t>$. We showed that if $f(t) \rightarrow f(0) $
as $t \rightarrow 0 $, the mean square-displacement after $n$
steps is proportional to $n/\ln(n)$. If $f(t) \sim f_{\alpha}
t^{-\alpha} $ ($0<\alpha<1$) as $t \rightarrow 0 $, we found that
the mean square-displacement is proportional to $
n^{(2-2\alpha)/(2-\alpha)}$. Quite interesting, we found that
anomalous diffusion of persistent walkers and {\it hopping}
particles on a site-disordered lattice\ \cite{alex,sahimi} are
similar. To observe photon subdiffusion experimentally, we
suggested a dielectric film stack for realization of a
distribution $f(t)$\ \cite{miriE}.


In the realm of diffusion on one-dimensional lattices with random
hopping rates $w_{\bm{j},\bm{j'}}$ from site $\bm{j'}$ to site
$\bm{j}$, the {\it asymmetric} hopping model with
$w_{\bm{j},\bm{j+1}} \neq w_{\bm{j+1},\bm{j}}$ has gained much
attention\ \cite{others1, others1a, others1b, others2, others3,
others4, others5}: At variance with the symmetric case, the
asymmetric model can display anomalous diffusion behavior without
broad distribution of hopping rates. The asymmetric hopping model
has been used to discuss hopping conductivity in presence of an
external electric field, molecular motors\ \cite{motors},
evolution of a domain wall in a one-dimensional random field Ising
model\ \cite{Bouchaud}, helix-coil transition of heteropolymers\
\cite{Bouchaud, dg}, etc. These points suggest us to investigate
{\it persistent} random walk of photons on a one-dimensional
lattice with random {\it asymmetric} transmittances. For a given
dielectric stack, the transmittance for incidence on the right
side, is equal to that for incidence on the left side\ \cite{yeh}.
However, optical elements with different transmission in the
forward and backward directions, and even {optical diodes} which
allow unidirectional propagation, are realized\ \cite{uni}. For
example, Fedetov et al.\ \cite{fedo} showed that asymmetric
transmission through a planar metal nanostructure consisting of
twisted elements can be observed in the optical part of the
spectrum. For a normally incident circularly polarized light of
wavelength $630~\mathrm{nm}$, this metamaterial is $1.3$ times
more transparent from one side than from the other. There is a
good reason to believe that the experimental observation of
photons' persistent random walk is not out of reach: Barthelemy,
Bertolotti, and Wiersma have recently verified {\it L\'{e}vy
flight} of photons in their synthesized L\'{e}vy
glass~\cite{nature}.

Apart from interest in the optics of random media, our work has
been motivated by the Lorentz gas model introduced to describe the
diffusion of conduction electrons in metals\ \cite{lo, hank}.
One-dimensional persistent random walk and {stochastic} Lorentz
gas are intimately related\ \cite{barkai2}. Stochastic Lorentz
model consists of fixed scatterers on a lattice and one moving
light particle. The light particle runs at velocity $c$ or $-c$,
and when collides with a scatterer it is transmitted with a
site-dependent probability $t$ or reflected with a probability
$1-t$. In other words, the light particle performs a persistent
random walk. Thus here we are investigating a variant of the
one-dimensional Lorentz gas, where each scatterer is characterized
by a random {\it asymmetric} transmission coefficient.

In this paper we consider the persistent random walk of photons on
a one-dimensional lattice of random asymmetric transmittances.
Each site is characterized by its intensity transmittance $t$
($t'$) for photons moving to the right (left) direction.
Transmittances at different sites are assumed independent,
distributed according to a given probability density
$\mathcal{F}(t,t')$. We generalize a variant of the effective
medium approximation introduced by Sahimi, Hughes, Scriven, and
Davis\ \cite{sahimi, sahimi2} to identify two classes of
$\mathcal{F}(t,t')$ which lead to the normal diffusion of photons:
(i) $ \langle {1}/{t}\rangle$ is finite and $ \langle
{t'}/{t}\rangle $ is less than $1$. (ii) $ \langle
{1}/{t'}\rangle$ is finite and $ \langle {t}/{t'}\rangle $ is less
than $1$. Here $\langle ~~\rangle $ denotes averaging with respect
to the distribution $\mathcal{F}(t,t')$.

Our paper is organized as follows. In Sec.\ \ref{model} we
introduce the model. In Sec.\ \ref{myema} we present the effective
medium approach to the problem. The numerical treatment and its
results are reported in Sec.\ \ref{montecarlo}. Sec.\
\ref{discuss} is devoted to a discussion of our results.

\section{Model}\label{model}

We consider a one-dimensional lattice random walk in which steps
are permitted to the nearest neighbor sites only. We normalize the
length and duration of a step to $1$. Apparently, on a
one-dimensional lattice the walker can move either to the right
($+$) or to the left ($-$) direction. Each site $\bm{j}$ is
characterized by forward and backward transmittances $
t_{\bm{j},\bm{j}+1} $ and $t'_{\bm{j},\bm{j}-1} $, respectively:
On arriving a site $\bm{j}$, a walker moving in the right (left)
direction takes a step in the same direction with the probability
$ t_{\bm{j},\bm{j}+1} $ ($t'_{\bm{j},\bm{j}-1} $). Here we assume
{\it asymmetric} transmittances, i.e. $ t_{\bm{j},\bm{j}+1} \neq
t'_{\bm{j},\bm{j}-1} $.

We assume that (i) transmittances $t$ and $t'$ at each site are
{random} variables. In general, these random variables are not
independent, (ii) transmittances at two different sites are
independent, (iii) transmittances at all sites are distributed
according to a given normalized probability density $
\mathcal{F}(t,t' )$. Apparently $\int_{0}^{1} \int_{0}^{1}
\mathcal{F}(t,t') dt dt'=1 $. The probability density functions of
$t$ and $t'$ are $ f^{+}(t)=\int_{0}^{1} \mathcal{F}(t,t') dt'$
and $f^{-}(t')= \int_{0}^{1} \mathcal{F}(t,t') dt$, respectively.
The joint probability distribution can be written as
$\mathcal{F}(t,t')=f^{+}(t) f^{-}(t')$ when random variables $t$
and $t'$ are independent. For any function $h(t,t')$, we define
$<h(t,t')>=\int_{0}^{1} \int_{0}^{1} h(t,t') \mathcal{F}(t,t') dt
dt'$.

We denote by $P^{+}(n, \bm{j})$ $\big ( P^{-}(n, \bm{j}) \big)$ the probability
that the walker after its $n$th step arrives at site $\bm{j}$ with positive (negative) momentum.
A set of two master equations can be established to couple the probabilities at step $n+1$
to the probabilities at step $n$:
\begin{eqnarray}
P^{+}(n+1, \bm{j})&=& t_{\bm{j}-1,\bm{j}}  P^{+}(n, \bm{j}-1) + r'_{\bm{j}-1,\bm{j}}  P^{-}(n, \bm{j}-1) ,\nonumber \\
P^{-}(n+1, \bm{j})&=&  r_{\bm{j}+1,\bm{j}} P^{+}(n, \bm{j}+1)  +t'_{\bm{j}+1,\bm{j}} P^{-}(n, \bm{j}+1) , \label{master1} \nonumber \\
\end{eqnarray}
where  $r_{\bm{j},\bm{j}-1} =1-t_{\bm{j},\bm{j}+1} $ and
$r'_{\bm{j},\bm{j}+1}  =1-t'_{\bm{j},\bm{j}-1}$ denote forward and
backward reflectances at site $\bm{j}$, respectively.

We are mainly interested in the probability that the photon
arrives at position $\bm{j}$ at step $n$, i.e. $ P(n,
\bm{j})=P^{+}(n, \bm{j})+P^{-}(n, \bm{j}) $, from which we extract
the first and second moments after $n$ steps as the characteristic
features of a random walk:
\begin{eqnarray}
 \langle \langle {\bm{j} }\rangle   \rangle _n  &=& \langle\sum_{\bm{j}}        \bm{j}    P(n, \bm{j})  \rangle ,\nonumber \\
\langle  \langle {\bm{j}^2 }\rangle \rangle _{n}   &=&\langle \sum_{\bm{j}}   \bm{j}^2 P(n, \bm{j}) \rangle .
\label{moments}
\end{eqnarray}
Here the first bracket represents an ensemble average over all random transmittances,
and the second bracket signifies an average with respect to the distribution $P(n, \bm{j})$.

Assuming a constant forward transmittance $t$ and a backward
transmittance $t'$ at each site, translational invariance of the
medium can be invoked to deduce the exact solution of $P(n,
\bm{j})$ in the framework of characteristic functions\
\cite{weiss}. Furthermore, the mean square-displacement of photons
after $n \rightarrow \infty $ steps can be obtained as $ \langle
{\bm{j}^2 } \rangle _{n}- \langle \bm{j} \rangle_{n}^{2}=2 D n$,
where the the diffusion constant $D$ is
\begin{equation}
D = \frac{2(1-t)(1-t')(t+t')}{(2-t-t')^3} , \label{hh}
\end{equation}
In the limit $t=t'$ one obtains $D=t/(2-2t) $, a known result in
the realm of the the classical persistent random walk.

The disorder not only may affect the value of diffusion constant
as compared to the ordered system, but also may lead to the
subdiffusive or superdiffusive behavior. In our model, even a few
sites with small transmittances may drastically hinder the photon
transport: In the extreme limit where at two different sites
$\bm{j}$ and $\bm{j'}$, transmittances $t_{\bm{j},\bm{j}+1} =
t'_{\bm{j},\bm{j}-1}= t_{\bm{j'},\bm{j'}+1} =
t'_{\bm{j'},\bm{j'}-1}=0$, photons either do not visit the segment
between $\bm{j}$ and $\bm{j'}$, or are caged in this segment. In
the following section, we determine which distributions of
transmittances $\mathcal{F}(t,t')$ lead to the normal diffusion of
photons.

\section{Effective Medium Approximation}\label{myema}

Many of the approaches to the transport in disordered media have
the disadvantage of being restricted to one-dimensional problems.
Here we adopt the effective medium approximation (EMA) which is
applicable to two- and three-dimensional media. We generalize a
variant of effective medium approximation introduced by Sahimi,
Hughes, Scriven, and Davis\ \cite{sahimi, sahimi2}.

First we simplify the set of coupled linear difference equations
(\ref{master1}) using the method of the $z$-transform\
\cite{weiss, jury} explained in Appendix\ \ref{zT}:
\begin{eqnarray}
\frac{P^{+}(z, \bm{j})}{z} -\frac{P^{+}(n=0, \bm{j})}{z}  &=&
t_{\bm{j}-1,\bm{j}}  P^{+}(z, \bm{j}-1)\nonumber \\
&& + r'_{\bm{j}-1,\bm{j}}  P^{-}(z, \bm{j}-1) ,\nonumber \\
\frac{P^{-}(z, \bm{j})}{z}- \frac{P^{-}(n=0, \bm{j})}{z}&=&  r_{\bm{j}+1,\bm{j}} P^{+}(z, \bm{j}+1)\nonumber \\
& &  +t'_{\bm{j}+1,\bm{j}} P^{-}(z, \bm{j}+1) . \label{master2}
\end{eqnarray}
We assume the initial conditions $P^{+}(n=0, \bm{j})=P^{-}(n=0,
\bm{j})=\delta_{\bm{j},0}/2$. To facilitate solution of Eq.
(\ref{master2}) we introduce probabilities $P^{\pm}_{e}(z,
\bm{j})$, and a reference lattice or average medium with all
forward transmittances equal to $t_{e} (z) $ and all backward
transmittances equal to $t'_{e} (z) $, so that
\begin{eqnarray}
\frac{P^{+}_{e}(z, \bm{j})}{z} -   \frac{P^{+}(n=0, \bm{j})}{z}&=&
t_{e} (z)  P^{+}_{e}(z, \bm{j}-1)\nonumber \\
& & + r'_{e} (z)   P^{-}_{e}(z, \bm{j}-1) ,\nonumber \\
 \frac{P^{-}_{e}(z, \bm{j})}{z}   - \frac{P^{-}(n=0, \bm{j})}{z}&=&
 r_{e} (z) P^{+}_{e}(z, \bm{j}+1) \nonumber \\
& & +t'_{e} (z)P^{-}_{e}(z, \bm{j}+1) . \label{masterema}
\end{eqnarray}
Here  $r_{e} (z)=1-t_{e} (z)$ and $r'_{e} (z)=1-t'_{e} (z)$ denote
effective reflectances.

EMA determines $t_{e} (z) $ and $ t'_{e} (z) $ in a
self-consistent manner, in which the role of distribution
$\mathcal{F}(t,t')$ is manifest. This is done by taking a cluster
of random transmittances from the original distribution, and
embedding it into the effective medium. We then require that {\it
average} of site occupation probabilities of the decorated medium
duplicate $P^{\pm}_{e}(z, \bm{j})$ of the effective medium. We
will sketch the method in the following.

Subtracting Eqs. (\ref{master2}) and (\ref{masterema}), we obtain
\begin{eqnarray}
&&\frac{1}{z}\!\!\begin{pmatrix}
\!Q^{+}(z,\!\bm{j}\!)\!\\\!Q^{-}(z,\!\bm{j}\!)\!\end{pmatrix}
\!\!-\!\mathbf{T}^{-}\!(z)\!\!\begin{pmatrix}
\!Q^{+}(z,\!\bm{j\!-\!1}\!)\!\\\!Q^{-}(z,\!\bm{j\!-\!1}\!)\!\end{pmatrix}
\!\!-\!\mathbf{T}^{+}\!(z)\!\!\begin{pmatrix} \!Q^{+}(z,\!\bm{j\!+\!1}\!)\!\\\!Q^{-}(z,\!\bm{\!j\!+\!1})\!\end{pmatrix} \nonumber \\
&&~~~= \Big [  \begin{pmatrix} t_{\bm{j-1},\bm{j}}  &
r'_{\bm{j-1},\bm{j}}\\ 0 & 0
\end{pmatrix} -\mathbf{T}^{-}(z)\Big]
\begin{pmatrix} P^{+}(z,\bm{j\!-\!1})\\P^{-}(z,\bm{j\!-\!1})\end{pmatrix}\nonumber \\
&&~~~~+\Big[  \begin{pmatrix}  0 & 0 \\ r_{\bm{j+1},\bm{j}} &
t'_{\bm{j+1},\bm{j}}
\end{pmatrix}-\mathbf{T}^{+}(z) \Big]
\begin{pmatrix} P^{+}(z,\bm{j\!+\!1})\\P^{-}(z,\bm{j\!+\!1})\end{pmatrix},
\label{mainema}
\end{eqnarray}
where
\begin{eqnarray}
\begin{pmatrix} Q^{+}(z,\bm{j})\\Q^{-}(z,\bm{j})\end{pmatrix} & =&
\begin{pmatrix} P^{+}(z,\bm{j})\\P^{-}(z,\bm{j})\end{pmatrix}
-\begin{pmatrix} P^{+}_{e}(z,\bm{j})\\P^{-}_{e}(z,\bm{j})\end{pmatrix},\nonumber \\
 \mathbf{T}^{-}(z) & =& \begin{pmatrix}  t_{e}(z) &r'_{e}(z) \\ 0 & 0 \end{pmatrix},\nonumber \\
 \mathbf{T}^{+}(z) & =& \begin{pmatrix}  0 & 0 \\ r_{e}(z) & t'_{e}(z) \end{pmatrix}.
\end{eqnarray}
To solve Eq. (\ref{mainema}), we introduce the Green function
$$\mathbf{G}(z,\bm{j})= \begin{pmatrix} G_{11} & G_{12} \\ G_{21} & G_{22}
\end{pmatrix}$$
which satisfies the equation
\begin{eqnarray}
\frac{1}{z} \mathbf{G}(z,\!\bm{j}\!) \!-\!\mathbf{T}^{-}\!(z)
\mathbf{G}(z,\!\bm{j}-1\!)
\!-\!\mathbf{T}^{+}\!(z)\mathbf{G}(z,\!\bm{j}+1\!)\! &=&\! \delta_{\bm{j},0} \mathbf{I}. \nonumber \\
\end{eqnarray}
Here $\mathbf{I}$ is the identity matrix. Multiplying both sides
of the above equation by $e^{\imath\bm{j}\theta}$ and then summing
over all the sites, the Fourier transform of the Green function
i.e. $ \mathbf{G}(z,\bm{\theta})=\sum_{\bm{j}=-\infty}^{\infty}
e^{\imath\bm{j}\theta} \mathbf{G}(z,\bm{j}) $, can be obtained as
\begin{eqnarray}
 \mathbf{G}(z,\bm{\theta})&=&\frac{z^2}{  \Delta(z,\bm{\theta})}
  \begin{pmatrix} \frac{1}{z}- {t'_{e}(z)} e^{-\imath\bm{\theta}} &  r'_{e}(z) e^{\imath\bm{\theta}} \\
 r_{e}(z) e^{-\imath\bm{\theta}} &  \frac{1}{z}- {t_{e}(z)} e^{\imath\bm{\theta}} \end{pmatrix}, \nonumber \\
\end{eqnarray}
where $  \Delta(z,\bm{\theta})=1-z
[{t_{e}(z)e^{\imath\bm{\theta}}+t'_{e}(z)e^{-\imath\bm{\theta}}}]
+ z^2[t_{e}(z)+t'_{e}(z) - 1]$.

For the present, we consider only the simplest approximation, and embed {\it one}
random transmittance at site $\bm{l}$ of the effective medium. Then solution of Eq. (\ref{mainema}) is
\begin{eqnarray}
\begin{pmatrix} Q^{+}(z,\bm{j})\\Q^{-}(z,\bm{j})\end{pmatrix} & = &
\int_{0}^{2 \pi} \mathbf{G}(z,\bm{\theta}) \mathbf{S}(z,\bm{\theta})  e^{-\imath\bm{\theta} (\bm{j} -\bm{l}) }\nonumber \\
& &~~~~~~\times \begin{pmatrix}
P^{+}(z,\bm{l})\\P^{-}(z,\bm{l})\end{pmatrix} \frac{d
\bm{\theta}}{ 2 \pi}, \label{poi}
\end{eqnarray}
where
\begin{eqnarray}
  \mathbf{S}(z,\bm{\theta}) \!=\!\!
 \begin{pmatrix} [t_{\bm{l},\bm{l+1}} -t_{e}(z)] e^{\imath\bm{\theta}} &  [t'_{e}(z)-t'_{\bm{l},\bm{l-1}} ]
 e^{\imath\bm{\theta}}\! \\
    [t_{e}(z)-t_{\bm{l},\bm{l+1}}] e^{-\imath\bm{\theta}}  & [t'_{\bm{l},\bm{l-1}} -t'_{e}(z)]
    e^{-\imath\bm{\theta}}\!\end{pmatrix}.
\end{eqnarray}

Self-consistency equation is $ <P^{\pm}(z,\bm{l})
>=P^{\pm}_{e}(z,\bm{l})$, or
\begin{equation}
< \big [\mathbf{I}-\int_{0}^{2 \pi} \mathbf{G}(z,\bm{\theta}) \mathbf{S}(z,\bm{\theta})
\frac{d \bm{\theta}}{ 2 \pi} \big]^{-1} >=\mathbf{I}.
\end{equation}
The above matrix equation leads to these conditions:
\begin{eqnarray}
& &\int_{0}^{1} \int_{0}^{1}  \frac{\mathcal{F}(t,t') dt dt'
}{1-(t-t_{e}(z)) U(z) -(t'-t'_{e}(z)) U'(z) } =1,\nonumber \\
& & \int_{0}^{1} \int_{0}^{1}  \frac{t  \mathcal{F}(t,t') dt dt'}{1-(t-t_{e}(z))U(z)-(t'-t'_{e}(z))U'(z)}
=t_{e}(z), \nonumber \\
& & \int_{0}^{1} \int_{0}^{1} \frac{t' \mathcal{F}(t,t') dt
dt'}{1-(t-t_{e}(z))U(z)-(t'-t'_{e}(z))U'(z)}=t'_{e}(z),\label{consist-eq3}
\nonumber \\
\end{eqnarray}
where
\begin{eqnarray}
U(z)&=& \frac{V(z)\big(1+z^2(- t_{e}(z) + t'_{e}(z) -1) \big) -1 }{ 2 t_{e}(z)} ,\nonumber \\
U'(z) &=&  \frac{V(z)\big(1+z^2(t_{e}(z) -t'_{e}(z) -1) \big) -1}{2 t'_{e}(z)} ,\nonumber \\
V(z)&=& \frac{1}{\sqrt{\big(1+z^2 (t_{e}(z)+t'_{e}(z)-1)\big)^2-4z^2t_{e}(z)t'_{e}(z)}}. \nonumber \\
\label{uv}
\end{eqnarray}
It turns out that one of the self-consistency conditions
(\ref{consist-eq3}) can be trivially satisfied. Consistency
equations determine $t_{e}(z)$ and $t'_{e}(z)$, in which the role
of distribution $\mathcal{F}(t,t')$ is manifest. For symmetric
transmittances where $\mathcal{F}(t,t')=f(t) \delta(t-t')$, our
consistency conditions (\ref{consist-eq3}) indeed duplicate that
of Ref.\ \cite{miriE}

The translational invariance of the effective medium can be
invoked to access the $z$-transform of the the first and second
moments of the photon distribution:
\begin{eqnarray}
\sum_{n=0}^{\infty}    \langle {\bm{j} } \rangle _{n} z^n &=&
\frac{z}{(1-z)^2}
\frac{t_{e}(z)-t'_{e}(z)}{1-z[t_{e}(z)+t'_{e}(z)-1]}, \nonumber  \\
\sum_{n=0}^{\infty} \langle {\bm{j}^2 }\rangle_{n}  z^n&=&
\frac{ 2 z^2}{(1-z)^3} \frac{(t_{e}(z)-t'_{e}(z))^2}{(1-z[t_{e}(z)+t'_{e}(z)-1])^2}\nonumber \\
&+&\frac{z}{(1-z)^2}
\frac{1+z[t_{e}(z)+t'_{e}(z)-1]}{1-z[t_{e}(z)+t'_{e}(z)-1]}.
\label{j2ema}
\end{eqnarray}
We are interested in the long time behavior, thus Tauberian
theorems suggest to analyze Eqs. (\ref{consist-eq3}) and
(\ref{j2ema}) in the limit $ z \rightarrow 1$.

To find which distributions of transmittances $\mathcal{F}(t,t')$
lead to the {\it normal diffusion} of photons, we assume that
$t_{e}(z)$ and $t'_{e}(z)$ have no singularity in the limit $ z
\rightarrow 1$. We find two distinct classes: (i) If $ \langle
{t'}/{t}\rangle  < 1 $ then
\begin{eqnarray}
t_{e}(z)  &=& \frac{1}{\langle {1}/{t}\rangle} , \nonumber  \\
 t'_{e}(z) &=&\frac{\langle {t'}/{t}\rangle}{\langle
{1}/{t}\rangle}. \label{gg1}
\end{eqnarray}
The first class of admissible distribution $\mathcal{F}(t,t')$ is
such that $ \langle {1}/{t}\rangle$ is finite and $ \langle
{t'}/{t}\rangle  $ is less than $1$. (ii) If $\langle
{t}/{t'}\rangle<1$ then
\begin{eqnarray}
t_{e}(z)  &=& \frac{\langle {t}/{t'}\rangle}{\langle
{1}/{t'}\rangle} ,\nonumber  \\
 t'_{e}(z) &=&\frac{1}{\langle
{1}/{t'}\rangle}. \label{gg2}
\end{eqnarray}
The second class of admissible distribution $\mathcal{F}(t,t')$ is
such that $ \langle {1}/{t'}\rangle$ is finite and $ \langle
{t}/{t'}\rangle  $ is less than $1$. The above effective
transmittances do not depend on $z$, thus one can directly use
Eq.~(\ref{hh}) to access the diffusion constant of photons.

The Cauchy-Schwarz inequality states that for two random variables
$ \alpha$ and $\beta $, $\langle  \alpha \beta \rangle^2 \leqslant
\langle  \alpha^2 \rangle \langle  \beta^2 \rangle  $. With
$\alpha=\sqrt{t/t'}$ and  $\beta=\sqrt{t'/t}$, we find that $ 1
\leqslant \langle {t}/{t'}\rangle   \langle {t'}/{t}\rangle $.
This clearly shows that $\langle {t}/{t'}\rangle$ and $\langle
{t'}/{t}\rangle$ are not simultaneously less than $1$, thus two
mentioned classes are quite distinct. EMA does not predict any
result when both $\langle {t}/{t'}\rangle$ and $\langle
{t'}/{t}\rangle$ are greater than $1$.

\section{Numerical simulations}\label{montecarlo}

\begin{figure}
\includegraphics[width=0.7\columnwidth]{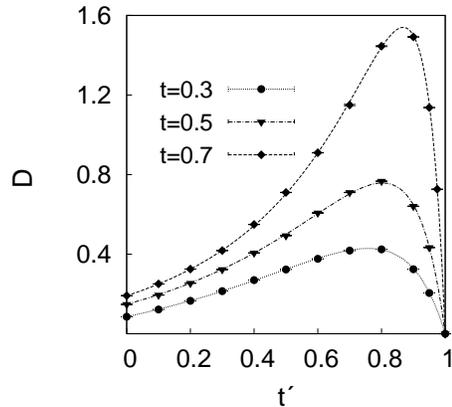}
\caption{ The diffusion constant $D$ as a function of
transmittances $t$ and $t'$ of an ordered medium. Theoretical and
Monte Carlo simulation results are denoted, respectively, by lines
and points.} \label{c1}
\end{figure}

\begin{figure}
\includegraphics[width=0.7\columnwidth]{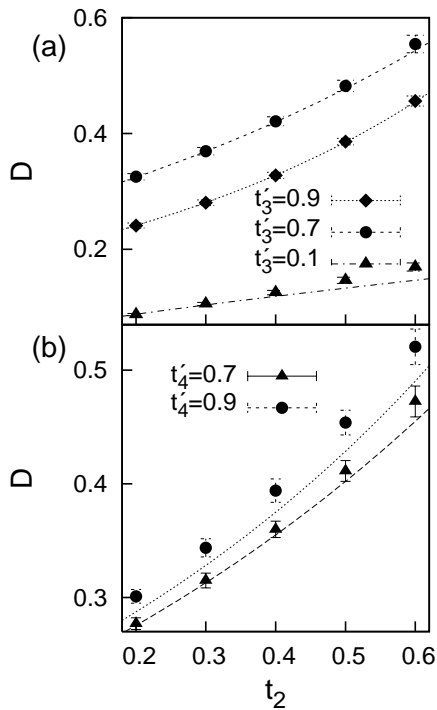}
\caption{(a) The diffusion constant $D$ as a function of $t_2$ for
different values of $t'_3$. $\mathcal{F}(t,t') =f^{+}(t)
f^{-}(t')$, such that $f^{+}(t)$ is a uniform distribution for
$0.2 <t < t_2$ and $f^{-}(t')= \delta(t'-t'_3)$. (b) $D$ as a
function of $t_2$ for different values of $t'_4$. $f^{+}(t)$ is a
uniform distribution for $0.1 <t <t_2$, and $f^{-}(t')$ is a
uniform distribution for $0.6 <t' <t'_4$. Theoretical and
simulation results are denoted, respectively, by line and points.}
\label{c2}
\end{figure}

The predictions of EMA can be inspected by numerical simulations.
The computer program produces $50$ media, whose transmittances are
distributed according to a given distribution $\mathcal{F}(t,t')$.
We deliberately focus on cases where both $\langle
{t}/{t'}\rangle$ and $\langle {t'}/{t}\rangle$ are not
simultaneously greater than $1$. For each medium, the program
takes $10^4$ photons at the initial position $\bm{j}=0$ and
generates the trajectory of each photon following a standard Monte
Carlo procedure. The statistics of the photon cloud is evaluated
at times $n \in [10000, 12000,...,68000]$. $\langle \langle
{\bm{j}^2 }\rangle \rangle _{n} - \langle \langle \bm{j} \rangle
\rangle_{n}^{2} $ is computed for each snapshot at time $n$, and
then fitted to $ 2 D n +O$ by the method of linear regression. An
offset $O$ takes into account the initial ballistic regime. We
compare our numerical diffusion constant with the analytical one
based on Eqs. (\ref{hh}), (\ref{gg1}) and (\ref{gg2}).

First we assume that transmittances of all sites are equal. For $
t \in [0.3, 0.5, 0.7]$ and $t' \in [0.0, 0.1, 0.2,...,0.9, 1.0]$,
our numerical and analytical predictions for the diffusion
constant are compared in Fig.~\ref{c1}.

Next we consider $\mathcal{F}(t,t')=f^{+}(t) f^{-}(t')$ such that
$f^{+}(t)$ is a uniform distribution for $t_1 <t < t_2$, and
$f^{-}(t')= \delta(t'-t'_3)$. We choose $t_1=0.2$, $0.2<t_2<0.6$,
and $t'_3 \in[0.1,0.7,0.9]$. Our numerical and analytical
predictions are compared in Fig.~\ref{c2}(a). We also considered
the case where $f^{+}(t)$ is a uniform distribution for $t_1 <t <
t_2$, and $f^{-}(t')$ is a uniform distribution for $t'_3 <t'<
t'_4$. We choose $t_1=0.1$, $0.2<t_2<0.6$, $t'_3=0.6$ and $t'_4
\in[0.7,0.9]$. Our results are shown in Fig.~\ref{c2}(b).

\begin{figure}
\includegraphics[width=0.7\columnwidth]{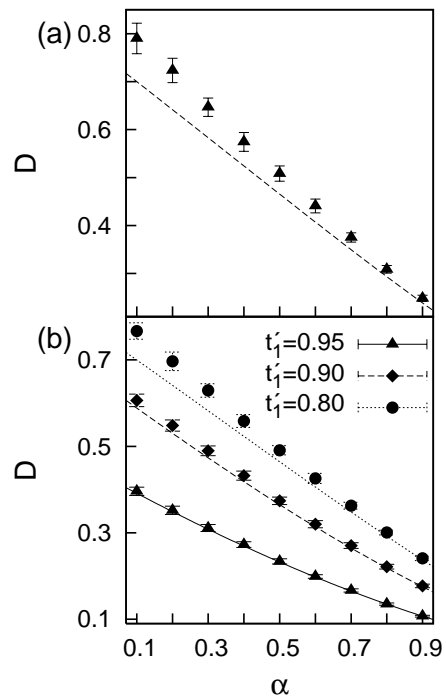}
\caption{The diffusion constant $D$ as a function of $\alpha$.
$\mathcal{F}(t,t')=f^{+}(t) f^{-}(t')$ such that (a) $f^{+}(t)
=(1-\alpha) t^{-\alpha} $ for $0 <t < 1$, and $f^{-}(t')$ is a
uniform distribution for $ 0.7 <t'<0.9 $. (b) $f^{+}(t)
=(1-\alpha) t^{-\alpha} $ and $f^{-}(t')=\delta(t'-t'_1)$.
Theoretical and simulation results are denoted, respectively, by
line and points.} \label{c3}
\end{figure}

We also present two other examples. We consider
$\mathcal{F}(t,t')=f^{+}(t) f^{-}(t')$ such that $f^{+}(t)
=(1-\alpha) t^{-\alpha} $ for $0 <t < 1$, and $f^{-}(t')$ is a
uniform distribution for $ 0.7 <t'<0.9 $. Fig.~\ref{c3}(a) depicts
$D$ as a function of $ \alpha$. Our results for the case $f^{+}(t)
=(1-\alpha) t^{-\alpha} $ and $f^{-}(t')=\delta(t'-t'_1)$ are
illustrated in Fig.~\ref{c3}(b).

Figures \ref{c1}-\ref{c3} vividly show that the effective medium
approach to the diffusion constant $D$ is quite successful.

\section{Discussions}\label{discuss}

In the present paper, we address the persistent random walk of
photons on a one-dimensional lattice of {random} asymmetric
transmittances. Clearly, a photon steps back by each reflection.
Intuitively, one expects the abundance of large reflectances (e.g.
{\it two} different sites with $t_{\bm{j},\bm{j}+1} =
t'_{\bm{j},\bm{j}-1}= t_{\bm{j'},\bm{j'}+1} =
t'_{\bm{j'},\bm{j'}-1}=0$) to drastically decrease excursion of
the photons. As percolation properties\ \cite{barry, sahimi1},
this feature is induced by the dimensionality of the lattice. We
focus on determining distributions of transmittances
$\mathcal{F}(t,t' )$ which lead to the normal diffusion of
photons. The probability distribution $P^{\pm}(n, \bm{j})$ as an
exact solution of the master equation (\ref{master1}) is quite
hard to obtain. However the relatively simple but approximate
effective medium approach reveals intriguing aspects of the
system. In two cases, the transport of photons is diffusive: (i) $
\langle {1}/{t}\rangle$ is finite and $ \langle {t'}/{t}\rangle $
is less than $1$. (ii) $ \langle {1}/{t'}\rangle$ is finite and $
\langle {t}/{t'}\rangle $ is less than $1$. Monte Carlo
simulations confirm our predictions. EMA does not predict any
result when both $\langle {t}/{t'}\rangle$ and $\langle
{t'}/{t}\rangle$ are greater than $1$.

It would be instructive to compare our problem with the transport
on a one-dimensional lattice with random asymmetric {\it hopping}
rates. Independent steps are the base of the hopping transport,
while correlated steps are the essence of the persistent random
walk. Hopping conduction is described by the master equation
\begin{eqnarray}
\frac{\partial P(\tau, \bm{j}) }{\partial \tau} &=& w_{\bm{j},
\bm{j}+1} P(\tau, \bm{j}+1)+ w_{\bm{j}, \bm{j}-1} P(\tau,
\bm{j}-1)\nonumber \\
 & &-(w_{\bm{j}+1, \bm{j}}+w_{\bm{j}-1,
\bm{j}}) P(\tau, \bm{j}), \label{exciton}
\end{eqnarray}
where  $P(\tau, \bm{j})$ is the probability for the particle to be
on site $\bm{j} $ at continuous time $\tau$, and
$w_{\bm{j},\bm{j'}}$ denotes the probability of jumping from site
$\bm{j'}$ to site $\bm{j}$ per unit time. In the {asymmetric}
hopping model $w_{\bm{j},\bm{j+1}} \neq w_{\bm{j+1},\bm{j}}$.
First we note that EMA does not predict any result when both
$\langle {w_{\bm{j},\bm{j}+1}}/w_{\bm{j}+1,\bm{j}}   \rangle =
\langle w_\leftarrow / w_\rightarrow \rangle $ and $\langle
{w_{\bm{j}+1,\bm{j}}}/w_{\bm{j},\bm{j}+1}   \rangle = \langle
w_\rightarrow/ w_\leftarrow \rangle$ are greater than $1$\
\cite{others2, others5}. Making use of a periodization of the
medium, Derrida obtained exact expressions for the velocity and
diffusion constant\ \cite{others1a}. In the case $ \langle
\log(w_\leftarrow / w_\rightarrow)  \rangle <0 $, he found (i) The
velocity $V$ vanishes if $ \langle w_\leftarrow / w_\rightarrow
\rangle \geqslant 1 $. (ii) For $ \langle w_\leftarrow /
w_\rightarrow \rangle  < 1 < \langle (w_\leftarrow /
w_\rightarrow)^2 \rangle $ the velocity is finite but the
diffusion coefficient is infinite. (iii) For $\langle
(w_\leftarrow / w_\rightarrow)^2  \rangle <1 $ both $V$ and $D$
are finite. All these results are easy to transpose when $ \langle
\log(w_\leftarrow / w_\rightarrow) \rangle > 0 $.

As already mentioned in Sec.~\ref{intro}, the metamaterial
introduced by Fedetov et al.\ \cite{fedo} can be used to realize a
lattice of random asymmetric transmittances. We suggest a simple
arrangement where a fraction $\varepsilon $ of the lattice sites
are randomly occupied by the metamaterial, and the rest of lattice
is occupied by half transparent dielectric slabs. One can measure
the diffusion constant of photons as a function of $\varepsilon$
to test our predictions. For the proposed distribution of
transmittances $\mathcal{F}(t,t' )=\varepsilon \delta(t-0.43)
\delta(t'-0.57) + (1-\varepsilon) \delta(t-0.5) \delta(t'-0.5)$,
we find
$$D=  \frac{ 2(1-0.2456 \varepsilon )}{(2-0.2456 \varepsilon)^2} .$$

EMA does not predict any result when both $\langle
{t}/{t'}\rangle$ and $\langle {t'}/{t}\rangle$ are greater than
$1$. It would be interesting to investigate the anomalous
diffusion of photons and self-averaging quantities of the system
following Refs.~\cite{others1a, others1b, others2}. Our studies
can also be extended to higher dimensional lattices.

\begin{acknowledgments}
We appreciate financial support from Iran Telecommunication
Research Center (ITRC).
\end{acknowledgments}

\appendix
\section{$z$-transform}\label{zT}
The $z$-transform $F(z)$ of a function $F(n)$ of a discrete variable $n=0, 1, 2,  ...$ is
defined by
\begin{equation}
{F}(z)=\sum_{n=0}^{\infty} F(n) z^n .
\label{ztrans}
\end{equation}
One then derives the $z$-transform of $F(n+1)$ simply as
$F(z)/z - F(n=0)/z$. Note the similarities of this rule with the
Laplace transform of the time derivative of a continuous function\ \cite{weiss, jury}.

Under specified conditions the singular behavior of $F(z)$ can be
used to determine the asymptotic behavior of $F(n)$ for large $n$
(Tauberian theorems)\ \cite{weiss}. For example, the identity $
{\Gamma(1-\alpha)} (1-z)^{\alpha-1}=\sum_{n=0}^\infty
\Gamma(n-\alpha+1){z^n}/{n!}$ shows that
\begin{eqnarray}
 F(z) \sim \frac{\Gamma(1-\alpha)}{(1-z)^{1-\alpha} } &\rightarrow& F(n) = \frac{\Gamma(n-\alpha+1)}{n!} ,
  \label{inv-laplas}
\end{eqnarray}
where $\Gamma(\alpha)= \int_{0}^{\infty} e^{-t} t^{\alpha-1} dt$.
Particularly,
\begin{eqnarray}
 F(z) \sim \frac{1}{(1-z)^{2}} &\rightarrow& F(n) = n+1 ,
 \nonumber \\
 F(z) \sim \frac{1}{(1-z)^{3}} &\rightarrow& F(n) =
 \frac{1}{2}(n^2+3n+2). \label{a3}
\end{eqnarray}

\end{document}